\documentstyle[12pt,a4]{article}
\newtheorem{theo}{Theorem}

\begin{document}
\title{The geometry of the Fisher selection dynamics}
\author{A. V. Shapovalov${}\sp a$ and E. V. Evdokimov${}\sp b$}
\date{}
\maketitle
{\it
${}\sp a$
Tomsk State University, Physics Department, 634050 Tomsk, Russia\\
e-mail: shpv@phys.tsu.ru
\par
${}\sp b$ Research Institute of Biology and Biophysics, 634050 Tomsk, Russia\\
e-mail: evd@biobase.tsu.tomsk.su}
\begin{abstract}
We study  the  Fisher model describing natural
selection in a population with a diploid structure
of a genome by differential- geometric methods.
For the selection dynamics we introduce an affine
connection which is shown to be the projectively
Euclidean and the equiaffine one. The selection
dynamics is reformulated similar to the motion
of an effective particle moving along
the geodesic lines in an 'effective external field'
of a tensor type. An exact solution is found to
the Fisher equations for the special case of fitness matrix
associated to the effect of chromosomal imprinting
of mammals. Biological sense of the differential- geometric
constructions is discussed. The affine curvature is considered
as a direct consequence of an allele coupling in
the system. This curving of the selection dynamics geometry is
related to an inhomogenity  of the time flow in the course
of the selection.
\end{abstract}
\vskip 10mm
{\bf Key words:} population dynamics, natural selection
\par
\noindent PACS: 87.10.+e

\section{Introduction}\label{Int}

The selection dynamics in biological populations was
usually investigated by the methods of the dynamic system theory.
Apart from simple cases of explicit integrability, the basic problem
in such an approach is to find  attractors of various types and
to study the stability problem \cite{Pykh}.
A survey of basic results of the several past
decades can be found in  monographs \cite{Poluect,Svir,Logofet}.
\par
In common practice of selection dynamics, the geometric methods
do not attract too much attention, although the importance
is stressed sometimes \cite{Svir,Hydro}. On the other hand,
these methods have demonstrated their efficiency in theoretical
and mathematical physics. In modern theoretical physics
we observe a trend to formulate dynamical principles
(having structural resemblance to the respective biological
relations) in terms of differential geometry and Lie algebra.
Geometric and algebraic methods provide both suitable mathematical
constructions bringing a system to integrability and they facilitate
the study of global characteristics of the system.
From this standpoint the geometric framework may have
considerable interest for the mathematical models
of biological systems.
\par
Population models of biological societies display a natural
hierarchy regarding to the degree and the character of
coupling between system components. In the most general form,
the population dynamics should be considered in the context
of the ecological system dynamics. The population enters
the system as a single element connected with the rest
ones by trophic, compete and other links. The system state
is characterized by its number, by genetic structure,
by age and sex distribution and some other quantities.
A complete description of the selection dynamics
(the change of the population genetic structure)
implies to consider  a multi-locus system with all the
genes  involved in the selection process.
To simplify the problem we may consider a single- locus
model. A classical example of such a case is
the well known Fisher model describing a coupling of alleles
of the given locus in the course of the selection.
The Darwin system model suggested by Eigen \cite{Eig79}
is an utmost case of this simplification.
It realizes the basic idea of the natural selection:
a differential survive of convariantly self-reduplicating units
(according to Timofeev-Resovskii nomenclature \cite{Tim}),
is realized in this model.
\par
In Ref. \cite{SE} we develop Hamiltonian form and thereby
simplectic geometric description for the selection
dynamics in the populations with haploid structure of a genome
(when any gene is represented in a single instance (single allele)).
\par
In the geometric framework the evolution of the Darwin system
with a stable organization is represented as a motion with
a constant velocity in an Euclidean plane space of
information variables. This is a result of absence of
a coupling between separate genotypes (quasispecies, by Eigen)
in the course of the selection.
\par
In the given work the differential- geometrical methods
are applied to the Fisher model which is one of the
upper level in the above hierarchy with respect to the Eigen model.
The Fisher model describes the natural selection in populations
with diploid structure of a genome. In this model, the necessary
condition for the reproduction (and therefore the selection)
of the individuals in the population, is a pairwise coupling
of separate haploid genotypes (gametes) by the zigote formation,
i.e. the cells or the organisms with the diploid (double)
gene setting. The gamete coupling in the population
is considered by the Fisher model in terms of pairwise coupling
of different alleles of the single gene (single-locus model)
or many genes (multi-locus model) being inherited by
the individual from its parents.
\par
The basic aim of the paper is to clarify the question
how the allele coupling creates the geometry
of the space associated with the selection dynamics.
We introduce an affine connection related to
the first derivative of the Fisher equations
in the population variables space. The affine connection
is turn out to be projectively Euclidean one and
simultaneously it is an equiaffine connection.
The above derivative is presented in the form of equations
of motion of an effective particle moving along the geodesic lines
in  an 'effective external field' of a tensor type.
The Fisher equations are integrated for the
fitness matrix of a special form associated with
the phenomenon of chromosomal imprinting and differentional
methylation of DNA in the course of the gamete maturation
of mammals \cite{Mark}. In this case the  Fisher dynamics
is reduced to the dynamics of the Darwin system by a
suitable replace of time variable. An approximate solution
is constructed in the weak 'external field'.
The equiaffinity is shown to result in the conservation
of a volume in the population space  with respect to
the geodesic flows. The volume is defined using a certain
density of the affine space.
\par
Mention that the curving of the Euclidean flat space
of informational variables is the effect of the allele
coupling in the system.
This curving could be also interpreted as an effect
of the inhomogenity of the time flow in the
in the course of the selection.

\section{ The Fisher model} \label{Fisher}

Consider a population where the success of reproductivity
(and therefore the selection dynamics) is determined by a single
gene (locus) having $N$ alleles.
Denote by  $p_\alpha $ a portion of $\alpha $-allele
in the population, $\alpha ,\beta ,\dots =1,\dots ,N$.
Evidently
\begin{equation}\label{1}
\sum _{\alpha  =1}^N p_\alpha =1.
\end{equation}
Choose the quantities  $p_\alpha $ as population variables.
Then the Fisher system is written as \cite{Fish}:
\begin{equation}\label{2}
\dot p_\alpha=p_\alpha (\sum _{\beta =1}^N\omega _{\alpha \beta }
p_\beta - \sum _{\mu, \nu =1}^N \omega _{\mu \nu} p_\mu p_\nu ).
\end{equation}
Here  $\dot p_\alpha=$ ${d p_\alpha (t)}/{dt}$,
$t$ is the time.
The quantity
$$
\sum _{\beta  =1}^N\omega  _{\alpha  \beta  }p_\beta
$$
is the specific rate of reproduction of ${\alpha}-$th
allele, and
$$
\sum _{\mu,  \nu =1}^N \omega_{\mu \nu} p_\mu p_\nu
$$
is an average specific rate of the population growth as a whole.
The matrix $\omega _{\alpha \beta}$ (= const) in the population biology
is identified to the fitness matrix associated with
separate alleles.
Let us take the $N-$th allele as a gauge one and introduce
the variables
\begin{equation}\label{3}
z^i=\ln \displaystyle \frac {p_i}{p_N}.
\end{equation}
Here and below $i,j,k, \dots =1,\dots ,N-1$.
The meaning of the variables $z^i$ is that
they  reflect  an information quantity per a degree of freedom
in the population in the frame of Shannon formalism
\cite{Shan} (we mean the information about the genetic
structure of the population).
\begin{theo} \label{theo1}
In the variables (\ref{3}) the Fisher system (\ref{2}) is reduced to the
form:
\begin{equation}\label{4}
\dot z^i=\alpha _i +
\displaystyle \frac {\sum _k \beta _{ik}\exp (z^k)}{1+\Omega}.
\end{equation}
\end{theo}
Here,
\begin{equation} \label{5}
\begin{array}{ll}
&\alpha _i=\omega _{iN}-\omega _{NN}, \quad \beta _{ik}=
\omega _{ik}-\omega _{Nk}-\omega _{iN}+\omega _{NN},\\
&\Omega =\sum _k \exp (z^k). \nonumber
\end{array}
\end{equation}
\par
{\bf Proof}. Introducing the auxiliary variables
$y_\alpha =\ln p_\alpha $, we have from (\ref{1}):
$\sum _\alpha \exp (y_\alpha )=1$. Then $\exp (y_N)=1-\sum _k\exp (y_k)$.
It is easy to verify that
\begin{eqnarray} \label{6}
&\exp (z^i)=\displaystyle {\frac {\exp (y_i)}{1-\sum _k \exp (y_k)}}, \quad
\exp (y_i)=\displaystyle {\frac { \exp (z^i)}{1+\Omega }},  \nonumber  \\
&\sum _k \exp (y_k)=\displaystyle \frac {\Omega}{1+\Omega}, \quad
\exp (y_N)=\displaystyle \frac {1}{1+\Omega }.
\end{eqnarray}
Eqs. (\ref{2}) in the variables  $y_\alpha$ take the form:
\begin{equation} \label{7}
\dot y_\alpha =\sum _\beta \omega _{\alpha \beta} \exp (y_\beta)
+\sum _{\mu ,\nu}\omega _{\mu \nu}\exp (y_\mu +y_\nu ).
\end{equation}
Writing Eqs. (\ref{7}) for $\alpha =i$ and for $\alpha =N$
and subtracting one from another, we obtain (\ref{4}) using
(\ref{6}).

\section{Projectively Euclidean space} \label{Geom}

Let us differentiate the system (\ref{4}) with respect to
$t$ and write down the result as follows:
\begin{equation} \label{8}
\begin{array}{ll}
&\ddot z^i =-\displaystyle \frac {1}{2(1+\Omega)}
\sum _{k,l}[\delta _{ik} \exp (z^l)+ \delta _{il}
\exp (z^k)]\dot z^k\dot z^l+\\
&\displaystyle \frac {1}{1+\Omega} \sum _k(\alpha _i+
\beta _{ik}) \exp (z^k)\dot z^k.
\end{array}
\end{equation}
Introduce the affine connection space
${\cal A}_{N-1}(\Gamma ^i_{kl})$.
In the coordinates  $z^i$ the connection is:
\begin{equation} \label{9}
\Gamma ^i_{kl} = \displaystyle \frac {1}{2(1+\Omega)}
[\delta _{ik} \exp (z^l)+ \delta _{il} \exp (z^k)].
\end{equation}
Eq. (\ref{8}) has the form of equations of motion for an
effective particle moving along the geodesic lines
in the affine connection space  ${\cal A}_{N-1}(\Gamma ^i_{kl})$
with the 'external field' defined by
the tensor field $A^i_k$:
\begin{equation} \label{10}
\ddot z^i + \sum_{kl}\Gamma ^i_{kl} \dot z^k \dot z^l =
\sum_{k}A^i_k \dot z^k,
\end{equation}
where,
\begin{equation} \label{11}
A^i_k= (\alpha _i+ \beta _{ik})
\displaystyle \frac {\partial \ln(1+\Omega)}{\partial z^k}.
\end{equation}
The original equations (\ref{4}) play the role of constraints
to Eqs. (\ref{10}).
\par
Consider the basic geometric properties of the space
${\cal A}_{N-1}(\Gamma ^i_{kl})$.
\par
The curvature tensor
\begin{equation} \label{12}
R_{lki.}^{\quad q} =
\displaystyle \frac {\partial \Gamma ^q_{li}}{\partial z^k}-
\displaystyle \frac {\partial \Gamma ^q_{ki}}{\partial z^l}+
\sum_{p}\Gamma ^q_{kp}\Gamma ^p_{li} -\sum_{p}\Gamma ^q_{lp}\Gamma ^p_{ki}
\end{equation}
can be written using (\ref{9}) as follows:
\begin{equation} \label{13}
\begin{array}{ll}
&R_{kql.}^{\quad j} = (1+\Omega)^{1/2}
[\displaystyle \frac {\partial ^2}{\partial z^k \partial z^l}
(1+\Omega)^{-1/2}\delta ^j_q -  \\
&\displaystyle \frac {\partial ^2}{\partial z^q \partial z^l}
(1+\Omega)^{-1/2}\delta ^j_k].
\end{array}
\end{equation}
The Ricci tensor, $R_{ql}=\sum_{k}R_{kql.}^{\quad k}$, is
\begin{equation} \label{13a}
R_{ql} = (1+\Omega)^{1/2} (2-N)
\displaystyle \frac {\partial ^2 (1+\Omega)^{-1/2}}
{\partial z^q \partial z^l}.
\end{equation}
Then, we can verify that
\begin{equation} \label{14}
R_{kql.}^{\quad j} = \frac {1}{2-N} (R_{kl}\delta ^j_q- R_{ql}\delta ^j_k).
\end{equation}
The case $N=2$ apparently requires special consideration. If
$N>2$, the curvature tensor satisfies the criterion of the
projectively Euclidean space \cite[p.540]{Rash}.
Hence, we get
\begin{theo} \label{theo2}
The space  ${\cal A}_{N-1}(\Gamma ^i_{kl})$  supplied with the
affine connection (\ref{9}) is a projectively Euclidean space.
\end{theo}
The form of the affine connection  $\Gamma ^i_{kl}$ (\ref{9})
leads to the statement that ${\cal A}_{N-1}(\Gamma ^i_{kl})$
is an equiaffine space \cite[\S41]{Norden}.
In such a space, a volume  exists  which is invariant under
the parallel transport of vectors.
The volume element spanned on vectors $\xi ^{i_k}_k$ has the
form
$$
V=\sum_{i_1,\dots ,i_{N-1} }e_{i_1\dots \i_{N-1}} \xi ^{i_1}_1
\dots \xi ^{i_{N-1}}_{N-1}
$$
which conserves under parallel transport of $\xi ^{i_k}_k$. Here
$e_{i_1\dots i_{N-1}} $ $=\sigma \epsilon _{i_1\dots i_{N-1}}$,
$\epsilon _{i_1\dots i_{N-1}}$ is completely antisymmetric symbol,
$\epsilon _{1\dots {N-1}}=1$.
In the space ${\cal A}_{N-1}(\Gamma ^i_{kl})$, $\sigma $
serves as a fundamental density and is defined by the following
condition
$$
\sum_{s}\Gamma ^s_{ks}=
\displaystyle \frac {\partial \ln \sigma }{\partial z^k}.
$$
For $\Gamma ^i_{kl}$ of the form (\ref{9}), we obtain:
$\sum_{s}\Gamma ^s_{ks}=$
$\partial \ln (1+\Omega)^{N/2}/\partial z^k$. Then, we have the
theorem true.
\begin{theo} \label{theo3}
The fundamental density of the space
${\cal A}_{N-1}(\Gamma ^i_{kl})$ with the connection
(\ref{9}) has the form  $\sigma =(1+\Omega)^{N/2}$.
\end{theo}
\par
The considered geometry is of a special interest
when the geodesical equations are exactly integrated.
In this case the allele coupling in our approach is
represented as a pure curving of the population
variable space.
It is connected with the following form of the fitness
matrix $\omega _{\alpha \beta}$ in the Fisher system (\ref{2}):
$\omega _{ij} = \omega _{Nj}, j\ne N$;
$\omega _{iN}\neq \omega _{jN}$, $i\neq j$.
As a variant of a real genetic system resulting in such form of
the matrix $\omega _{\alpha \beta}$ we can suggest
a system with the chromosomal imprinting
found in the course of the gametogenesis
of mammals.  According to \cite{Mark}, the result of the
imprinting is that the same allele is included into a genome
of a zigote in active or in non-active state. The allele activity
depends on its affiliation to the DNA of male or female gamete.
Molecular mechanism of the chromosomal imprinting is based on the
differential methylation of DNA in the course of the
gamete maturation.
\par
For  weak  'external field' $A^i_k$, that takes place under
the condition
\begin{equation} \label{15}
\alpha _i+\beta _{ik}=\varepsilon \gamma _{ik},\quad  \varepsilon \ll 1,
\end{equation}
($\alpha _i$, $\beta _{ik}$ are of the form (\ref{5}))
the analysis of the motion can be carried out in two steps.
First, we consider the motion along the geodesic lines, then we
account for the 'external field' $A^i_k$ perturbatively.

\section{ Integration of the geodesic equations} \label{Geodesic}

The system (\ref{10}) has the form of equations of geodesics
of the space
${\cal A}_{N-1}(\Gamma ^i_{kl})$ under the condition $A^i_k=0$.
Eqs. (\ref{10}), with account of (\ref{9}), can be written in the form:
\begin{equation}\label{16}
\ddot z^j+\dot z^j
\sum _l\displaystyle \frac {\partial \ln (1+\Omega)}{\partial z^l} \dot z^l=0,
\end{equation}
or
$$
\ddot z^j+\dot z^j
\displaystyle \frac {d }{d t}\ln (1+\Omega) =0.
$$
The first integration gives
\begin{equation} \label{17}
\dot z^j = \displaystyle \frac {c_j }{1+\Omega},
\end{equation}
where $c_j-$ are integration constants.
\par
Notice that  the original equations (\ref{4})
under the condition (\ref{15}) take the form:
\begin{equation} \label{18}
\dot z^j = \displaystyle \frac {\alpha _j }{1+\Omega}+
\displaystyle \frac {\varepsilon \sum _k \gamma _{jk} \exp (z^k)}{1+\Omega}.
\end{equation}
Putting $\varepsilon =0$, we see that the restrictions being imposed
by Eqs. (\ref{4}) on the system (\ref{16}), are reduced to
\begin{equation} \label{19}
c_j = \alpha _j.
\end{equation}
Let us now integrate Eqs. (\ref{17}) under the condition (\ref{19}).
Without loss of generality, we can assume
\begin{equation} \label{20}
z^j = \alpha _j \varphi (t) + b_j.
\end{equation}
Here $b_j$ are the integration constants, and the function
$\varphi (t)$ is determined by the equation
$$
\dot \varphi (t)=\displaystyle \frac {1}{1+\Omega},\quad
\Omega =\sum _k \exp (\alpha _k\varphi (t) +b_k),
$$
whose implicit solution is
\begin{equation} \label{21}
\varphi (t)+\sum _k \alpha ^{-1}_k \exp (\alpha _k\varphi (t)+b_k)=t-t_0.
\end{equation}
Let us redefine the parameter in the equations  of integral
lines (\ref{20}) setting $\tau= \varphi (t)$.
Then the time $t$ is explicitly expressed in terms of the
parameter $\tau$ as
\begin{equation} \label{22}
t-t_0= \tau +\sum _k \alpha ^{-1}_k \exp (\alpha _k\tau +b_k).
\end{equation}
Eqs. (\ref{20}) take the form of equations of the motion
with the constant velocity  $\alpha _j$ with respect
to the 'new time' $\tau$ for an effective particle:
\begin{equation} \label{23}
z^j = \alpha _j \tau + b_j.
\end{equation}
For a weak  'external field' $A^j_k$ (\ref{15}), Eqs. (\ref{10})
can be solved approximately as follows.
\par
Let us put
\begin{equation}\label{24}
z^j=\alpha _j \tau +b_j (\tau)
\end{equation}
and take $b_j$ to be a slowly varying function of $\tau$.
Substituting (\ref{24}) into (\ref{18}) and taking into account
(\ref{19}), we get:
\begin{equation}\label{25}
\displaystyle \frac {d b_j}{d \tau} =
\varepsilon \sum_k \gamma _{js} \exp (\alpha _s\tau +b_s).
\end{equation}
In the first approximation (putting $b_s$ to be constants
in the right-hand side of (\ref{25})) we obtain:
\begin{equation}\label{26}
z^j=\alpha _j\tau +
\varepsilon \sum _k \gamma _{jk} \alpha ^{-1}_k \exp (\alpha _k+b_k(0)),
\end{equation}
where  $b_k(0)$ are the constants of integration.

\section{ Conclusion} \label{Discuss}

Discuss a biological sence  of the parameters
$\alpha _i$, $\beta _{ij}$  in (\ref{4}).
If  $\beta _{ij}=0$ then the system (\ref{4})
is reduced to the dynamics of
haploid populations. A detail analysis of such systems
was performed in  \cite{SE}.
The parameters  $\alpha _i$ in this case have a meaning of
specific rates of reproduction of separate genotypes
(the Maltusian parameters in ecological terms).
Thus, it is the matrix  $\beta _{ij}$
(not the complete matrix $\omega _{\alpha \beta}$) which
is effectively responsible  for the alleles coupling. It is not
evident from the original Fisher system (\ref{2}).
The allele coupling is represented as a curvature in
the geometry of the population variable space.
The tensor (\ref{12}) is the measure of this curvature
and it is completely determined by the value $\ln (1+\Omega)$.
Note that this value is expressed in the population variables
in the following form: $\ln (1+\Omega)=\ln (1/p_N)$.
In its turn,  $\ln (1/p_N)$ is a measure of an information
of the $N-$th allele in the population by Shannon.
It is clear, the more rarely the $N-$th allele occurs
in the population the larger is the curvature of the space
associated with the selection dynamics.
Let us also emphasize that the basic geometrical
characteristic of this space is completely determined
by encounter frequency of $N$-th allele.
Among all the possible variants of the matrices
$\omega _{\alpha \beta}$, a particular interest
has the case $\varepsilon=0$ in (\ref{15})
when the system (\ref{4}) is integrated explicitly
and its solution is reduced to the uniform rectilinear
motion (\ref{23})  by the redefinition of time variable.
This case is  similar to the Eigen selection dynamics
in haploid populations and, as it is shown in  \cite{SE},
admits the Hamiltonian form of the dynamic equations.
Let us note that in this case the allele coupling
is effectively reduces to time flow change according to
(\ref{22}).
\par
The work was supported by RFFR grant 98-02-16195.
\par

\end{document}